\documentclass{article}

    \PassOptionsToPackage{numbers, compress}{natbib}

\usepackage{amsmath, amssymb, tabularx}  
\usepackage{graphicx}          
\usepackage{booktabs}

\usepackage{enumitem}
\usepackage{microtype}
\usepackage{wrapfig}
\usepackage{adjustbox}

    \usepackage[preprint]{neurips_2025}

\usepackage[utf8]{inputenc} 
\usepackage[T1]{fontenc}    
\usepackage{hyperref}       
\usepackage{url}            
\usepackage{booktabs}       
\usepackage{amsfonts}       
\usepackage{nicefrac}       
\usepackage{microtype}      
\usepackage{xcolor}         

\title{SQL-of-Thought: Multi-agentic Text-to-SQL with Guided Error Correction }

\author{%
  Saumya Chaturvedi \\
  Max Planck Institute for Software Systems\\
  Saarbrücken, Germany \\
  \texttt{schaturv@mpi-sws.org} \\
  \And
  Aman Chadha \\
  AWS GenAI \\
  Santa Clara, CA, USA \\
  \texttt{hi@aman.ai} \\
  \And
  Laurent Bindschaedler \\
  Max Planck Institute for Software Systems \\
  Saarbrücken, Germany \\
  \texttt{bindsch@mpi-sws.org} \\
}

\begin{document}

\maketitle

\begin{abstract}
Converting natural language queries into SQL queries is a crucial challenge in both industry and academia, aiming to increase access to databases and large-scale applications. This work examines how in-context learning and chain-of-thought can be utilized to develop a robust solution for text-to-SQL systems. We propose SQL-of-Thought: a multi-agent framework that decomposes the Text2SQL task into schema linking, subproblem identification, query plan generation, SQL generation, and a guided correction loop. Unlike prior systems that rely only on execution-based static correction, we introduce taxonomy-guided dynamic error modification informed by in-context learning. SQL-of-Thought achieves state-of-the-art results on the Spider dataset and its variants, combining guided error taxonomy with reasoning-based query planning.
\end{abstract}

\section{Introduction}
Text-to-SQL (NL2SQL) has emerged as a crucial problem in both research and real-world applications, enabling non-technical users to query structured databases through natural language. While early sequence-to-sequence and schema-aware models improved accessibility, they often struggled with generalization, resulting in subpar performance. Recent advances in large language models (LLMs) and prompting techniques have significantly raised performance, with methods such as DIN-SQL \cite{dinsql} and DAIL-SQL \cite{dailsql} decomposing the task into subtasks. However, as highlighted by Li et al. \cite{dawnofnl2sql} and Biswal et al. \cite{tag}, even the most advanced systems remain brittle when dealing with realistic queries. This is because execution-based feedback alone cannot correct logically incorrect, yet syntactically valid, SQL queries.

Multi-agent approaches \cite{toolsql, chasesql, naacl-multiagent} and reasoning-guided prompting \cite{exploringcot, think2sql} have shown potential to bridge this gap. Multi-agent systems enhance modularity and specialization; however, their error correction still relies heavily on execution signals or static re-generation. On the other hand, reasoning-based methods improve query planning, yet unguided reasoning can often introduce new errors or inefficiencies. What is needed is a systematic way to combine multi-agent decomposition with structured reasoning and interpretable correction.

We tackle this issue using SQL-of-Thought, a multi-agent framework designed to enhance SQL queries. This framework incorporates several key components: (i) specialized agents for schema linking, identifying subproblems, generating query plans, synthesizing SQL, and correcting errors; (ii) a taxonomy-guided error correction loop that categorizes SQL failure modes and rectifies them through a Chain of Thought reasoning process; and (iii) an analysis comparing reasoning and non-reasoning models across different agent roles. Our framework achieves state-of-the-art execution accuracy on Spider \cite{spider}, Spider-Realistic \cite{spider-realistic}, and Spider-SYN \cite{spider-syn}, while also showcasing effective cost-performance trade-offs through the use of hybrid models. These results demonstrate that compact, guided reasoning, paired with structured error feedback, is a more reliable approach for SQL querying than solely relying on execution-based refinement.

We make the following contributions:
\begin{itemize}
    \item SQL-of-Thought: A novel multi-agent solution for NL2SQL, complete with Schema Linking, Subproblem identification, Chain of Thought (CoT) for query plan generation, SQL generation, and an explicit error correction loop.
    \item Guided Error Modification driven by a comprehensive Error Taxonomy and In-Context Learning
    \item A detailed analysis of Reasoning and Non-reasoning model choices for SQL-of-Thought
    \item State of the Art results on Spider \cite{spider} benchmark and its variations: Spider-Realistic \cite{spider-realistic}, and Spider-SYN \cite{spider-syn}.
\end{itemize}

\begin{figure}
    \centering
    \includegraphics[width=\linewidth]{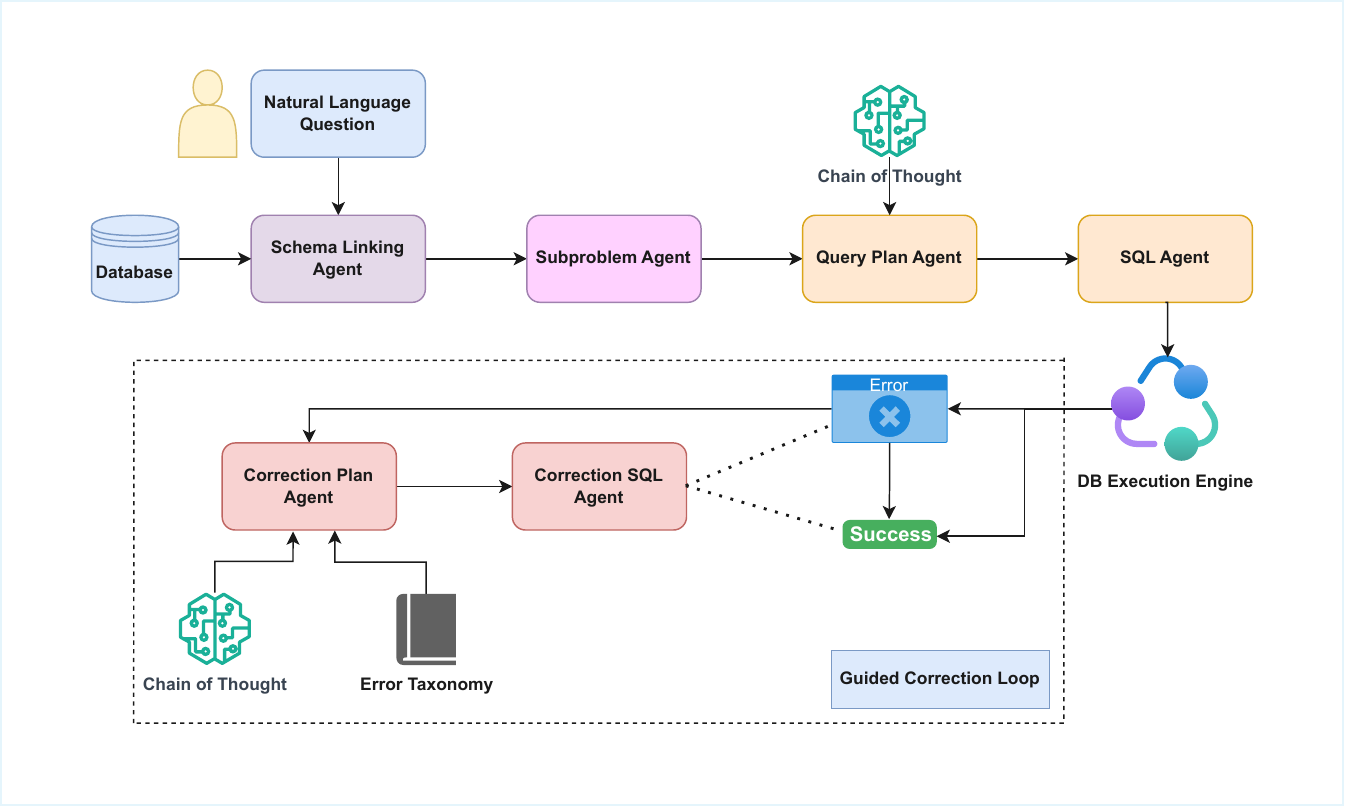}
    \caption{Architecture of SQL-of-Thought. The process begins as the user asks a natural language question. The relevant schema is fetched from the database, and sent alongwith the question to Schema Linking agent, which extracts the useful tables and columns needed to solve the question. This cropped schema and the question is sent to the Subproblem agent, which divides the problem into smaller, clause-wise subproblems. Then the Query Plan Agent is invoked, using Chain of Thought to procedurally generate an executable Query Plan. This query plan is used by the SQL Agent to generate a valid SQL solution, which is then executed on the DB Execution engine. If the execution fails, the correction loop is invoked, with the execution error, error taxonomy, and incorrect SQL sent to the Correction Plan Agent, to investigate the error and generate a Correction Plan using a CoT-guided rectification mechanism to fix it. The Correction SQL Agent uses this plan to generate the correct SQL query. The Guided Correction Loop is invoked until the query is successful or maximum attempts are finished.}
    \label{fig:architecture}
\end{figure}

\section{Related Work}
\subsection{History of NL2SQL}
Initial works for Text-to-SQL used major sequence-to-sequence architectures such as PLMs (Programming Language Models), for example RESDSQL \citep{resdsql}, which identified relevant schema from the question and then used it to construct the SQL query. Other choices were GNNs (Graph Neural Networks), Recurrent Neural Networks (RNNs), Transformers, and most recently LLMs for natural language to SQL conversion. With the advent of LLMs, prompt engineering on powerful models like GPT \citep{gpt3, gpt4} and Claude enabled fine-grained and faster solutions for NL2SQL. Shi et al \citep{llm-survey} highlights how LLM-based solutions outperformed all other alternatives for programming text-to-SQL. 

DIN-SQL \citep{dinsql} attempted in-context learning with error correction, decomposing subproblems, and defining different prompt templates to tackle each subtask. However, their method only regenerates prompts in case of failure, without specific error information. DAIL-SQL \cite{dailsql} designs sophisticated SQL-code style prompt templates to encode the question and database schema. These templates provide few-shot examples for schema linking and SQL query generation, utilizing GPT-4 to generate solutions. DAILSQL-SC employs self-consistency to postprocess solutions for further refining.

Some studies \cite{dawnofnl2sql} discuss production readiness and evaluation frameworks for NL2SQL using various prompting methods, again verifying that LLM-based approaches generally outperform pretrained task-specific models through higher reliability and generalization. They classify queries into four categories: subqueries, joins, logical connectors, and order-by clauses, to measure model efficacy across increasing levels of difficulty. The TAG framework \cite{tag} showed how existing models fail on complex real-world queries, highlighting that most NL2SQL models and benchmarks are only effective for roughly 20\% of realistic user queries. They introduce a reasoning framework that decomposes the problem into three stages: (i) query synthesis through NL2SQL parsing, (ii) query execution over the underlying database, and (iii) answer generation, where an LLM integrates both the natural language query and execution results to produce a final human-readable answer.

The TAG framework \cite{tag} demonstrated how existing models struggle with complex real-world queries and proposed steps for query synthesis, execution, and answer generation. Multi-agent frameworks, such as Wang et al. \cite{macsql}, introduced syntax validation and execution optimization, but relied mainly on execution feedback.

\subsection{Agentic Approaches}
Tool-SQL \citep{toolsql} pioneered the tool-first agentic framework, where an LLM-based agent iteratively generates and refines SQL queries with the aid of two tools: a database retriever to resolve condition mismatches and an error detector for stricter constraint validation. The approach primarily addressed database mismatches, such as missing highlights, altered paraphrasing, and similar issues.

Chase SQL \citep{chasesql} defines an elaborate framework consisting of value retrieval, multiple candidate generators and query fixers, and a selection agent. The candidates generate a query plan by prompting the LLM through an "EXPLAIN" keyword. The multi-agent framework proposed by Shao et al. \citep{naacl-multiagent} introduces a four-agent system where specialized roles (developer, researcher, executor, specialist) collaborate on simple and complex NL2SQL queries. By combining example retrieval, chain-of-thought reasoning, syntax validation, and error-driven refinement, the framework improves both correctness and execution efficiency.

\subsection{Reasoning guided solutions to NL2SQL}
ACT-SQL \cite{actsql} experiments with prompting styles and few-shot exemplar selection strategies to automate prompt selection for natural language to SQL solutions, making the process cheap and time-saving. Think2SQL \citep{think2sql} and related methods explored reasoning for text-to-SQL by combining zero-shot prompting, supervised finetuning, and reinforcement learning. They asked whether reasoning improves text-to-SQL, how best to train reasoning-capable models, and whether execution-based rewards are sufficient. Their results show mixed evidence: reasoning models often perform better on query planning but not consistently across datasets. They also simplified schema linking by manually providing subsets, limiting the scope of evaluation. Tai et al. \citep{exploringcot} highlight Chain-of-Thought prompting techniques, which achieve significant gains on the Spider dataset, and mention how unnecessarily detailed reasoning can propagate errors.

Our work differs in that we employ a chain of thought reasoning approach, not only for planning but also for schema linking, subproblem division and error correction, guided by an explicit taxonomy. We also evaluate our approach on a variety of models, both reasoning and general-purpose LLMs, to verify the importance of model CoT strength.

\section{Methodology}
The proposed SQL-of-Thought framework is implemented as an LLM-driven text-to-SQL pipeline that generates an executable SQL query $Y$. The process can be formalized as:
$$Y = LLM(Q, S, C, P, T | \theta)$$

where $Q$ is the natural language question, $S$ is the linked schema, $C$ denotes clause-specific subproblems, $P$ is the query plan and $T$ represents the error taxonomy used for CoT-informed error correction. $LLM(.|\theta)$ is a language model parameterized by $\theta$.

\subsection{SQL-of-Thought}

A detailed composition of the SQL-of-Thought framework is as follows:

\textbf{Schema Linking Agent} The Schema Linking Agent parses the natural language question in conjunction with the database schema (identified via \textit{db\_id}) to identify the relevant tables and columns required for answering the query. In addition, it extracts structural information such as primary keys, foreign keys, and join relationships. This representation forms the foundation for subsequent steps by constraining SQL generation to schema-relevant entities.

\textbf{Subproblem Agent}. Given the natural language question and the schema-linked output, the Subproblem Agent decomposes the query into clause-level subproblems (e.g., WHERE, GROUP BY, JOIN, DISTINCT, ORDER BY, HAVING, EXCEPT, LIMIT, UNION). Each identified clause is expressed as a key–value pair in a structured JSON object, where the key is the clause type and the value is the partially completed clause expression. This decomposition provides a modular representation of the query intent, enabling downstream agents to reason over smaller, well-defined units.

\textbf{Query Plan Agent}. The Query Plan Agent generates a step-by-step execution plan that maps the user’s intent to the schema and subproblems. Unlike prior work such as Chase-SQL \cite{chasesql}, which treats planning as a surface-level mapping, our agent is explicitly prompted to perform chain-of-thought reasoning to explain intermediate decisions. This structured reasoning encourages deeper alignment between the question, subproblems, and the final SQL query. The Query Plan Agent produces only the procedural plan and is explicitly restricted from generating executable SQL at this stage.

\textbf{SQL Agent}. The SQL Agent consumes the natural language question and the query plan to generate the executable SQL query. Post-processing removes extraneous artifacts such as trailing semicolons or natural language fragments, ensuring the query is syntactically valid. The generated query is then executed against the database, and the result is compared with the ground-truth answer. If the query fails to match, the pipeline transitions to the correction loop.

\textbf{Correction Plan Agent}: The Correction Plan Agent initiates the correction loop by analyzing the failed SQL query in the context of the natural language question, schema, and execution results. Unlike systems such as DIN-SQL or DAIL-SQL \cite{dailsql, dinsql}, which rely solely on execution feedback, this agent is additionally guided by an error taxonomy derived from Shen et al. \cite{icl-errors}. The taxonomy categorizes common error modes (e.g., schema mismatches, join inconsistencies, aggregation misuse), and the agent produces a chain-of-thought plan describing how to resolve the identified errors. Inspired by reflexive learning approaches \cite{reflexion}, the agent iteratively generates structured feedback that informs the correction SQL generation process.

\textbf{Correction SQL Agent}: The Correction SQL Agent takes as input the correction plan, the question, the schema, and the incorrect SQL query. Based on the structured guidance, it regenerates the SQL query while avoiding the previous errors. The corrected query is re-executed on the database, and if the result still does not match, the correction loop is re-entered until convergence or until the maximum number of correction attempts is reached.

\begin{figure}
    \centering
    \includegraphics[width=\linewidth]{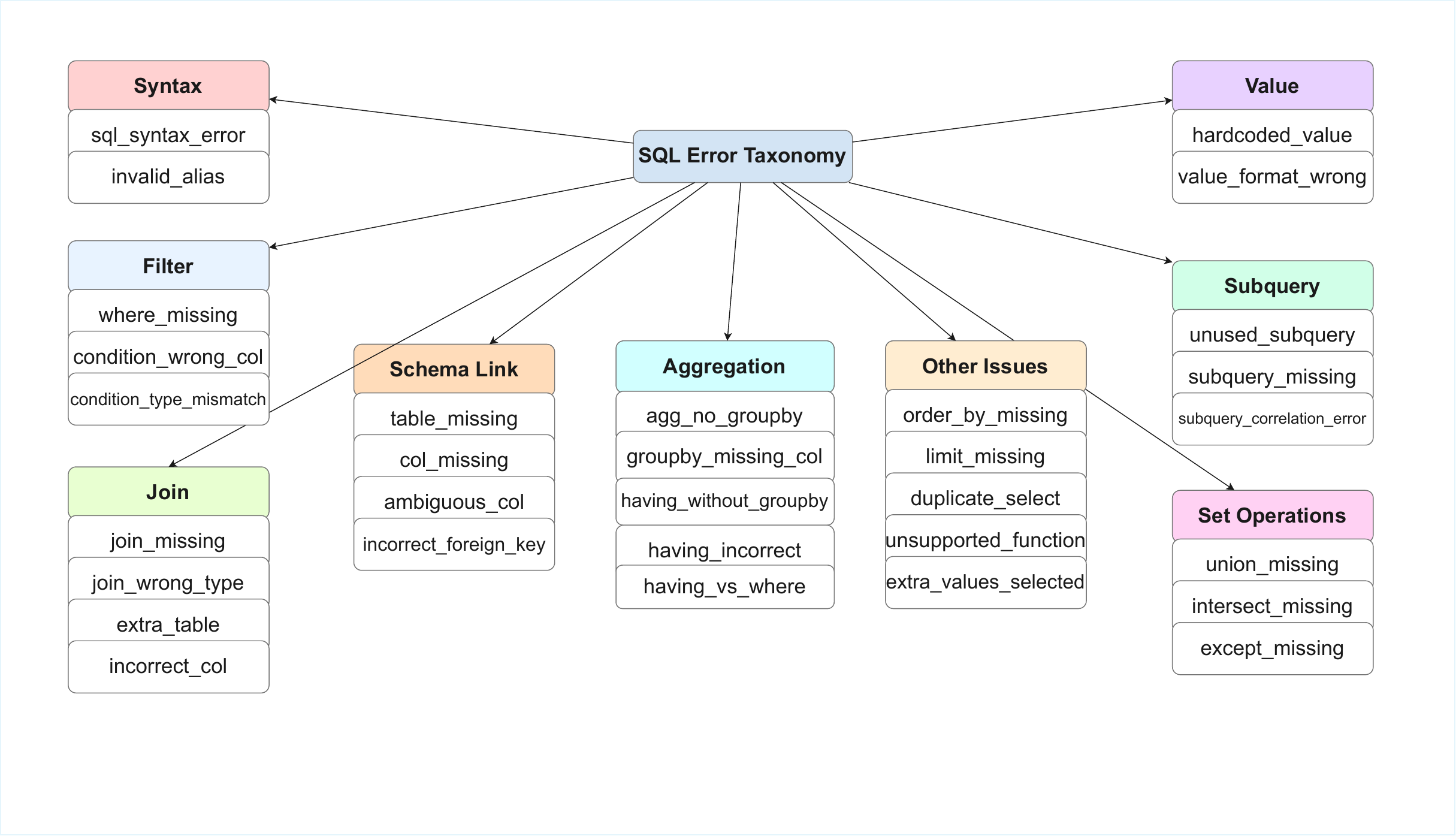}
    \caption{Error Taxonomy proposed for SQL-of-Thought. This taxonomy has 9 categories and 31 sub-categories of logical errors to be identified and rectified by LLMs.}
    \label{fig:error-taxonomy}
\end{figure}

\subsection{Error Taxonomy}
The error taxonomy adopted in our correction loop, shown in Figure \ref{fig:error-taxonomy}, is derived from the classification proposed in \citet{icl-errors}, which systematically categorizes standard failure modes encountered in NL2SQL systems. Our proposed taxonomy expands on this by creating systematic categories of errors that can be identified and corrected by LLMs, as well as comprehensive subtypes for each. We provide concise error codes instead of lengthy explanations to facilitate easier identification and prevent overflowing the LLM context window. Our taxonomy spans a broad range of issues, including \textbf{syntax errors} (e.g., invalid aliases or malformed SQL), \textbf{schema linking errors} (e.g., missing or ambiguous columns, incorrect foreign keys), and \textbf{join-related mistakes} (such as missing joins, wrong join types, or inclusion of extra tables). It further covers errors in \textbf{filter} conditions (e.g., incorrect columns in the WHERE clause, type mismatches), \textbf{aggregation logic} (e.g., missing GROUP BY, misuse of HAVING), as well as \textbf{value representation errors} like hard-coded values or format mismatches. More advanced categories capture failures in \textbf{subquery} formulation (e.g., unused or incorrectly correlated subqueries), \textbf{set operations} (union, intersection, except), and other structural oversights, such as missing ORDER BY or LIMIT clauses. By explicitly codifying these error types, the taxonomy provides a fine-grained diagnostic lens that enables the correction plan agent to reason not just about what failed, but why it failed within the SQL generation pipeline, ranging from basic syntax issues to complex sub-query and join formulation errors, providing specific guidance for correction.

Incorporating this taxonomy into the correction loop enables the system to move beyond coarse execution-based feedback (as used in DIN-SQL \cite{dinsql} and DAIL-SQL \cite{dailsql}) and instead provide interpretable, linguistically grounded guidance for iterative refinement. We prompt the correction plan agent with the summarized error taxonomy, along with a chain-of-thought reasoning template, which equips the agent to identify root causes of failure, align errors with schema or syntax constraints, and propose concrete repair strategies. This approach aligns with findings in reflexive learning approaches \citep{reflexion}, where LLMs benefit from verbalized self-feedback in enhancing decision-making for programming tasks. Thus, by leveraging the structured taxonomy from \citet{icl-errors}, the multi-agent system establishes a principled loop of error detection, diagnosis, and guided correction, thereby closing the gap between raw execution feedback and systematic, explainable query refinement.

\subsection{Experimental Setup}

\subsubsection{Dataset}
We utilize the Spider dataset \cite{spider} for evaluating our multi-agent framework. Spider offers a diverse range of database schemas and tasks, including 20 database settings and 1034 text-SQL pairs in its dev split. We avoid the BIRD-SQL benchmark due to the high quantity of annotation errors, as reported in Shen et al \cite{icl-errors}. BIRD-SQL has fewer set operation-based questions than SPIDER, and also presents an evidence field for queries, which can add to hallucination while query generation. For these purposes, we test on Spider and its popular variant, Spider Realistic \cite{spider-realistic}, which is an evaluation set based on the Spider dev set with explicit mentions of column names removed, containing 508 samples. Spider SYN \cite{spider-syn} is a challenging
variant of the Spider evaluation dataset. Spider- SYN is constructed by manually modifying natural language questions with synonym substitutions. 

\subsubsection{Evaluation Metrics}
While we observe metrics such as Exact Match and Valid SQL Generation during our experiments, we only consider Execution Accuracy for holistic evaluation. One natural language question could have multiple SQL queries that answer it. LLMs frequently oversimplify, and in the case of Text-to-SQL tasks, they were found to assign variables to results from subqueries, which is why an Exact Match (EM) is not an accurate representation of the accuracy of the process. Thus, we adopt Execution Accuracy (EA) as the metric to evaluate our process. EA is a boolean field that compares the execution results of the generated query with the execution results of the Gold SQL label that exists in the dataset. Since the EA calculation requires the generated SQL query to be executed against Spider's SQLite files, we also adopt SQLite \cite{sqlite} for query generation.

\subsubsection{Hardware \& Configuration} 

We perform all experiments on a machine equipped with two NVIDIA H100 GPUs, each with 80 GB of HBM, and configured with Debian Ubuntu, running PyTorch version 2.5.1.

\subsection{Guided Error modification}
Execution-based feedback alone is insufficient for error correction in NL2SQL. In our experiments, we observed that when using models such as Claude 3 Opus or GPT-4o-mini within the SQL-of-Thought framework, 95–99\% of generated queries were already syntactically valid. Thus, traditional database execution errors were rare, and the predominant failures stemmed instead from \textit{intent mismatches}, logically incorrect but syntactically valid queries, or errors involving incorrect or extraneous values through various join and limit clauses. Providing only execution traces \cite{dinsql, dailsql} therefore offers little guidance for effective correction. To address this, we integrate the error taxonomy defined in Figure \ref{fig:error-taxonomy}, which enumerates a wide range of SQL error types. By exposing the model to these structured error categories, the correction loop enables the agent to identify better and rectify subtle logical and semantic flaws. Empirically, on our strongest configuration (Claude 3 Opus + SQL-of-Thought), omitting the correction loop led to a 10\% increase in incorrect queries compared to the whole system with guided error modification, as demonstrated through ablation on a 100-sample evaluation set in Table \ref{different-model-results}.  

\subsection{Query Plan and then SQL Generation}
We also find that reasoning-driven LLMs perform significantly better when first producing a structured query plan before attempting SQL generation. The query plan serves as an intermediate reasoning step, allowing the model to explicitly organize relevant schema elements, logical conditions, aggregations, and subproblems before constructing the final query. This structured decomposition reduces hallucinations and improves alignment between the natural language intent and the SQL output. In our ablation study with Claude 3 Opus + SQL-of-Thought, bypassing the query plan agent and directing the SQL agent to generate queries directly increased the rate of incorrect queries by 5\% on a 100-sample test set, shown in Table \ref{different-model-results}. These results highlight the importance of staged reasoning—query plan synthesis followed by SQL generation as a critical design choice for improving correctness in NL2SQL systems.  


\section{Results}
We present the results of the proposed SQL-of-Thought framework on the Spider \cite{spider} and Spider Realistic \cite{spider-realistic} benchmarks (where results were available) in Table \ref{tab:combined_results}, highlighting the execution accuracy reported by earlier works and demonstrating how SQL-of-Thought outperforms existing methods. For all our experiments, we set the temperature to 0, and set top\_p and other hyperparameters to default. These results are based on the Claude Opus 3 model, which serves as the foundation for evaluating our SQL-of-Thought framework. SQL-of-Thought reports an Execution Accuracy of \textbf{91.59\%} on Spider \cite{spider} Benchmark and \textbf{90.16\%} Execution Accuracy on Spider Realistic \cite{spider-realistic} benchmark. We also report an EA of \textbf{82.01\%} on Spider SYN \cite{spider-syn} benchmark. We also wish to report that for all our experiments, we observe a Valid SQL generation rate of 94\%-99\% with SQL-of-Thought.

\vspace{1mm}
\begin{table}[ht]
\centering
\begin{tabular}{lcc}
\toprule
\textbf{Method} & \textbf{Spider} & \textbf{Spider-Realistic} \\
\toprule
ChatGPT \cite{gpt3} & 74.4 & - \\
GPT-4 (OpenAI 2023) \cite{gpt4} & 72.3 & - \\
ACT-SQL + ChatGPT (Zhang et al. 2023) \cite{actsql} & 80.4 & 75.8 \\
ACT-SQL + GPT-4 (Zhang et al. 2023) \cite{actsql} & 82.9 & - \\
DIN-SQL + GPT-4 (Pourreza and Rafiei 2024) \cite{dinsql} & 82.8 & 78.1 \\
DAIL-SQL + GPT-4 (Gao et al. 2023) \cite{dailsql} & 83.1 & 75.6 \\
DAIL-SQL + GPT-4 + SC (Gao et al. 2023) \cite{dailsql} & 83.6 & 75.2 \\
MAC-SQL + GPT-4 (Wang et al. 2024) \cite{macsql} & 86.8 & - \\
Tool-SQL + GPT-4 \citep{toolsql} & 86.9 & 82.9 \\
Chase SQL \citep{chasesql} & 87.6 & - \\
\textbf{SQL-of-Thought (with Claude Opus 3)} &  \textbf{91.59} & \textbf{90.16} \\
\bottomrule
\end{tabular}
  \vspace{2mm}
\caption{Execution Accuracy of prior methods and our proposed SQL-of-Thought on Spider (Dev split) and Spider-Realistic (Dev split). Dashes indicate that the corresponding result was not reported by the authors.}
\label{tab:combined_results}
\end{table}
\vspace{1mm}

Table \ref{tab:different-model-results} denotes the results obtained when different models were employed as base LLMs in SQL-of-Thought on a 100-sample mini-batch from the Spider Dev dataset. We also evaluated the framework once with just the initial sequential flow, removing the self-correction loop, and once with direct SQL generation after Schema Linking and Subproblem Agents, removing the CoT Query Plan generation during both the initial and correction flows. The accuracy falls at least 5\% without a Chain-of-Thought Query Plan before the SQL generation, and at least 8-10\% for all models when the correction loop is not invoked. The models are listed in decreasing order of accuracy, noting that Claude Opus 3 performs the best, with GPT-5, GPT-4o-mini \cite{gpt4}, and finally GPT-3.5 \cite{gpt3} producing \textbf{89\%, 87\%} and \textbf{67\%} accuracy, respectively.

We also report that a single run on Spider Dataset for SQL-of-Thought with Claude Opus 3, took 5 hours to run, and \$42.58 token cost at \$15/MTok (million tokens).

\begin{table}[ht]
  \centering
  \label{tab:different-model-results}
  \setlength{\tabcolsep}{8pt} 
  \begin{tabularx}{\textwidth}{lccc} 
    \toprule
    \textbf{Model} & {SQL-of-Thought} & {W/o Error Correction} & {W/o Query Plan Generation} \\
    \midrule
    \textbf{Claude 3 Opus} & \textbf{95} & \textbf{85} & \textbf{90} \\
    {GPT 5}         & 89 & 85 & 88 \\
    {GPT 4o Mini}   & 87 & {72} & {79} \\
    {GPT 3.5}       & {67} & {59} & {73} \\
    \bottomrule
  \end{tabularx}
  \vspace{1mm}
    \caption{Execution Accuracy Results for the Multiagent SQL model on the first 100 samples of Spider Dataset, evaluated using various base models and removal of correction loop, and query plan generation step to highlight the results.}
\end{table}

\section{Discussions}
\subsection{Reasoning vs. Non-Reasoning Models}
Reasoning models demonstrated clear advantages in nearly all specialized agentic tasks, including schema linking, query planning, chain of thought, and clause identification. Non-reasoning models often failed to decompose queries correctly and produced valid but logically wrong SQL. Errors such as confusing ColumnA from TableA and listing it under ColumnB, forgetting to include aggregate groupby clauses, selecting extraneous or fewer columns, and more were prevalent. However, reasoning alone did not guarantee improvements in all cases. For example, unguided reasoning without an error taxonomy led to worse correction and often resulted in the repetition of the same correction steps in different attempts. This acutely highlights how the dynamic guided correction loop complemented the Chain-of-Thought style prompting and helped the model learn from direct examples and an error playbook.

\subsection{Learnings}
Using Claude Opus 3 instead of GPT improved accuracy. The guided correction loop, where the agent relied on an error taxonomy, reduced repetition and improved logical corrections. Reasoning-based query plan generation before SQL synthesis also consistently improved results, as LLMs have been shown to be better at formulating reasoning for mathematical, programming, and planning tasks. In our ablation study with Claude 3 Opus + SQL-of-Thought, bypassing the query plan agent and directing the SQL agent to generate queries directly increased the rate of incorrect queries by 5\% on a 100-sample test set, shown in Table \ref{tab:different-model-results}. These results highlight the importance of staged reasoning: query plan synthesis followed by SQL generation as a critical design choice for improving correctness in NL2SQL systems.  

\subsection{Lessons from Unsuccessful Ablations}
Not all design choices improved accuracy; several variants underperformed, highlighting the importance of guided reasoning, as mentioned below:
\begin{itemize}
    
\item A critic loop that applied the full error taxonomy in a free-form way and sent errors directly to the SQL agent underperformed. Accuracy improved when errors were first routed through a query plan agent, showing that LLMs struggle with unguided debugging and benefit from at least one structured reasoning step.

\item Increasing temperature above 0 for GPT-4o added surface diversity but reduced plan faithfulness, producing more invalid joins and clause misuse.

\item Adding specific rules in the agent prompts for specific clauses (e.g., JOIN, LIMIT) inflated the context window and often distracted the model with irrelevant details, lowering accuracy.

\item An ablation design with multiple repair agents, each addressing a specific error type of a subproblem and feeding into an aggregation agent to generate the final SQL solution, failed. Independent edits often conflicted, and the merge process produced incoherent SQL.

\item  Carrying history across correction attempts through a shared scratchpad expanded the context window, increased latency and API cost, and amplified repetition and schema drift, leading to lower accuracy.

\end{itemize}

\subsection{Token Cost and other Metrics}
LLMs are better and faster at NL2SQL \cite{llm-survey, dawnofnl2sql}, but the API costs are very expensive. Especially in a multi-agent framework or any multiple-pipeline system, the cost for each sample is a lot. Increasing accuracy comes at the cost of compute and expensive processes. A single run on the Spider benchmark takes about 42.58\$ for Claude Opus 3, and 44.2\$ for GPT models. Switching to cheaper models led to a dip in accuracy, but good-performing models were more expensive. We tried to find a middle ground that is both cost-effective and high-performing. With various combinations of Claude and GPT model inference for different agents, we discovered that tasks such as Schema Linking, Query Plan Generation, and Correction Plan Generation required higher reasoning power and failed when using non-reasoning models. This meant that the Subproblem, SQL, and Correction SQL Agents could use non-reasoning models and still perform at par. This approach enabled a hybrid model of Claude Opus for reasoning-intensive agents and GPT-4o for the other agents to create a good middle ground at approximately a \$30 cost for an entire dataset run, achieving 85\% Execution Accuracy (for a 100-sample ablation).

Additionally, we also tested SQL-of-Thought on open-source models, such as LLama-3.1-8B-Instruct \citep{llama3} and Qwen2.5-1.5B \citep{qwen2.5}, and similar variations. We found that not only do these models exhibit high latency, taking up to three times longer for evaluation, but they also underperform significantly, achieving only about 45.3\% accuracy on the 100-sample set. These models had trouble generating SQL (led to long strings of repeated hallucination), missing column names across different tables, and were overall not suited for use in an NL2SQL framework. Future work can improve the cost and efficacy of such a framework by fine-tuning SLMs (small language models) on individual agentic tasks. For example, an error database can be designed to fine-tune the Correction Loop agents. Spider \cite{spider} provides detailed fields for each query, including whether it uses clauses such as LIMIT, INTERSECT, GROUPBY, etc. These clauses can be used to create datasets for fine-tuning the Subproblem Agent.

\section{Limitations}
Our evaluation is limited to the Spider \cite{spider} benchmark and its variants, which may not fully capture the challenges of real-world databases. The error taxonomy, although effective, has not been exhaustively validated across diverse query structures. The multi-agent design also increases inference costs, relying heavily on closed-source reasoning LLMs. Future work can improve both cost and efficacy by fine-tuning small language models (SLMs) for individual agentic tasks, such as building an error database to fine-tune correction loop agents or utilizing Spider’s clause-level annotations (e.g., LIMIT, INTERSECT, GROUPBY) to fine-tune the Subproblem Agent.

\section{Conclusion}
We introduced SQL-of-Thought, a multi-agent NL2SQL framework that combines schema linking, subproblem identification, chain-of-thought query planning, SQL generation, and taxonomy-guided error correction. The framework achieves state-of-the-art execution accuracy on Spider benchmarks, showing that structured reasoning and guided correction are more effective than execution-only feedback. Future work should extend evaluation to real-world datasets and explore fine-tuned SLMs for cost-efficient deployment.

\bibliographystyle{plainnat}
\bibliography{neurips_2025}

\clearpage
\newpage
\appendix
\renewcommand{\thesubsection}{A\arabic{subsection}}

\end{document}